\documentclass{article}

\usepackage[table]{xcolor} %
\usepackage{subcaption}
\usepackage{microtype}
\usepackage{graphicx}
\usepackage{booktabs} %
\usepackage{layouts}
\usepackage{hyperref}
\usepackage{xspace}
\usepackage[most]{tcolorbox}
\usepackage{makecell}
\usepackage{multirow}

\usepackage[accepted]{icml2025}

\usepackage{amsmath}
\usepackage{amssymb}
\usepackage{mathtools}
\usepackage{amsthm}

\usepackage[capitalize,noabbrev]{cleveref}

\theoremstyle{plain}

\theoremstyle{definition}

\theoremstyle{remark}

\crefname{section}{Section}{Sections}
\Crefname{section}{Section}{Sections}
\crefname{figure}{Figure}{Figures}
\Crefname{figure}{Figure}{Figures}
\crefname{subfigure}{Figure}{Figures}
\Crefname{subfigure}{Figure}{Figures}
\crefrangelabelformat{subfigure}{#3#1#4--#5(\crefstripprefix{#1}{#2}#6}

 \newcommand{\highlight}[1]{\textcolor{red}{#1}}

\newcommand{\system}{M1-Parallel\xspace}
 
\icmltitlerunning{Optimizing Sequential Multi-Step Tasks with Parallel LLM Agents}

\begin{document}

\twocolumn[
\icmltitle{Optimizing Sequential Multi-Step Tasks with Parallel LLM Agents}

\icmlsetsymbol{intern}{*}

\begin{icmlauthorlist}
\icmlauthor{Enhao Zhang}{uw,intern}
\icmlauthor{Erkang (Eric) Zhu}{msr}
\icmlauthor{Gagan Bansal}{msr}
\icmlauthor{Adam Fourney}{msr}
\icmlauthor{Hussein Mozannar}{msr}
\icmlauthor{Jack Gerrits}{msr}
\end{icmlauthorlist}

\icmlaffiliation{uw}{University of Washington, Seattle, USA}
\icmlaffiliation{msr}{Microsoft Research, Redmond, USA}

\icmlcorrespondingauthor{Erkang (Eric) Zhu}{erkang.zhu@microsoft.com}

\icmlkeywords{Machine Learning, ICML}

\vskip 0.3in
]

\printAffiliationsAndNotice{\textsuperscript{*}Research performed during an internship at Microsoft }  %

\begin{abstract}
Large language model (LLM)-based multi-agent systems have demonstrated remarkable promise for tackling complex tasks by breaking them down into subtasks that are iteratively planned, executed, observed, and refined. 
Despite their effectiveness, these systems often incur high latency because real-world problems frequently demand multiple iterative cycles of reasoning steps.
To address this challenge, we propose \system, a framework that concurrently runs multiple multi-agent teams in parallel to uncover distinct solution paths. By leveraging an event-driven communication model with asynchronous messaging, \system efficiently capitalizes on the inherent diversity of valid plans to either reduce end-to-end latency or boost task completion rates.
Our experiments on complex tasks show that \system with early termination achieves up to $2.2\times$ speedup while preserving accuracy, and that \system with aggregation yields higher task completion rates. We further investigate strategies aimed at encouraging diverse execution plans but observe no additional performance gains over repeated sampling. 
Overall, these findings underscore the potential of parallel plan execution for optimizing multi-agent systems for real-world, high-complexity reasoning tasks.
\end{abstract}

\newcommand{\TaskFigure}{
    \begin{figure*}[t!]
        \centering
        \includegraphics[width=0.95\textwidth]{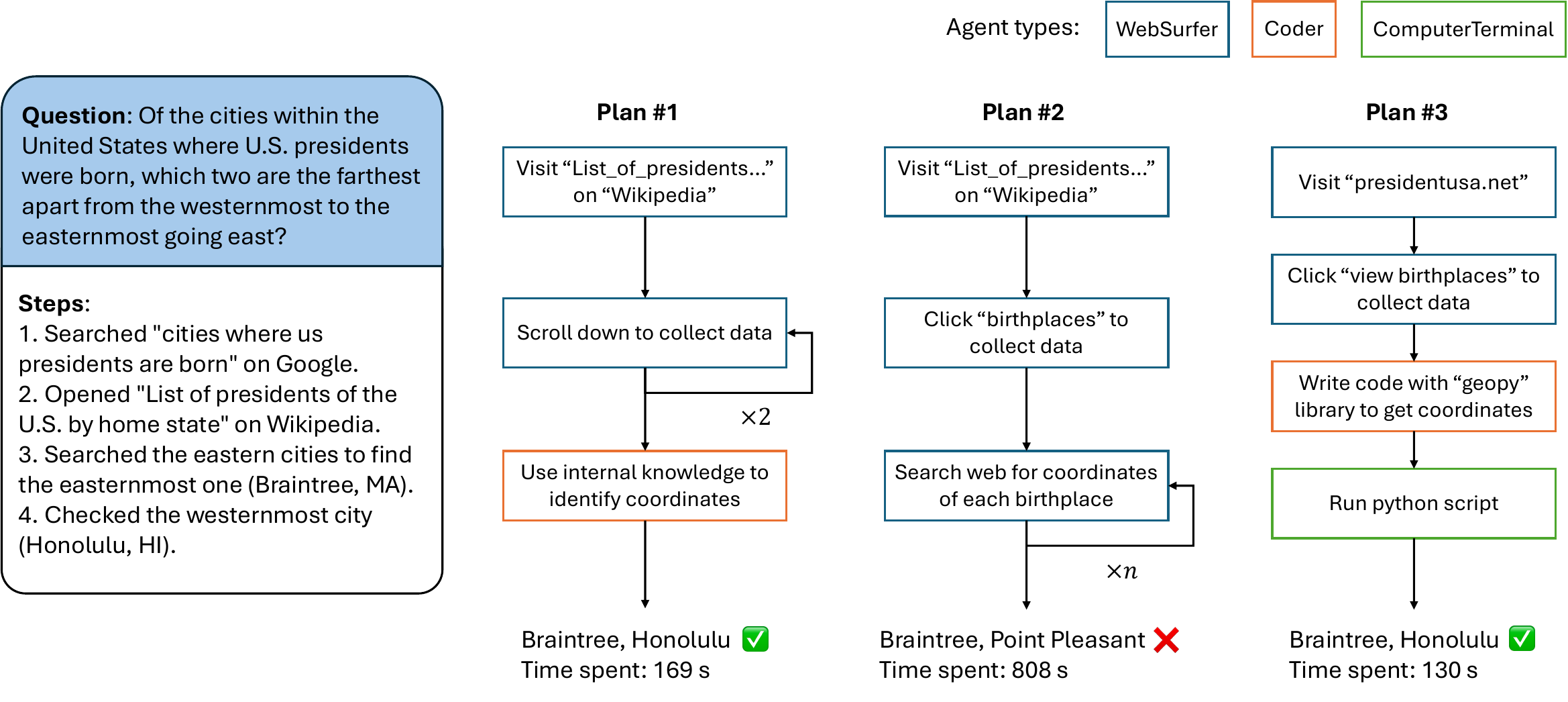}
        \caption{Left: An example task with annotated solution steps from GAIA. Right: Multiple plans for solving the same task. \system reduces end-to-end latency by executing multiple plans in parallel, noting that a complex task often has multiple valid solving plans with different latencies. In this example, \system launches three plans in parallel and terminates when the execution of the fastest plan (i.e., Plan \#3) finishes. Plan \#2 yields an incorrect answer because the LLM agents decide to end the task early after detecting an lengthy, repetitive pattern in searching coordinates of birthplaces.}
        \label{fig:example_task}
    \end{figure*}
}

\newcommand{\SystemDiagramFigure}{
    \begin{figure}[t!]
        \centering
        \begin{subfigure}{0.49\columnwidth}
            \centering
            \includegraphics[width=\columnwidth]{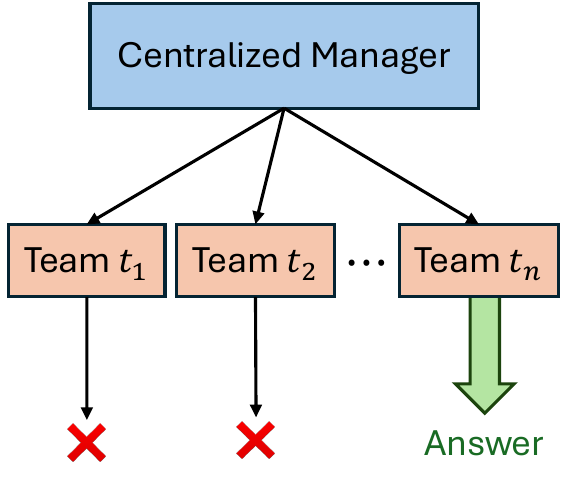}
            \caption{Early stopping.}
            \label{subfig:early_stop}
        \end{subfigure}
        \begin{subfigure}{0.49\columnwidth}
            \centering
            \includegraphics[width=\columnwidth]{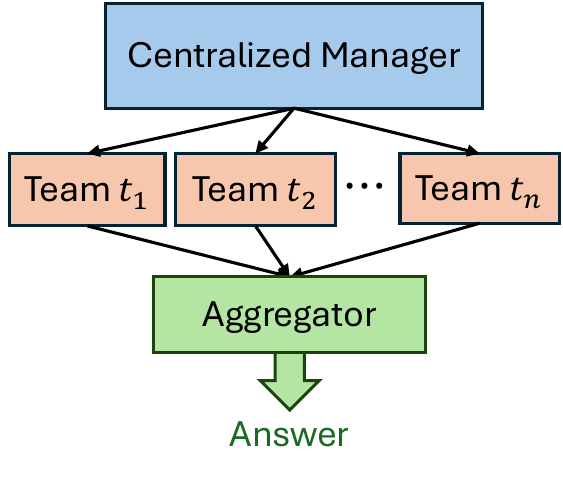}
            \caption{Aggregation.}
            \label{subfig:aggregation}
        \end{subfigure}
        
        \caption{Parallel agents with (a) early stopping and (b) aggregation.}
        \label{fig:system_diagram}
    \end{figure}
}

\newcommand{\GaiaMainFigure}{
    \begin{figure}[t!]
        \centering
        \includegraphics[width=\columnwidth]{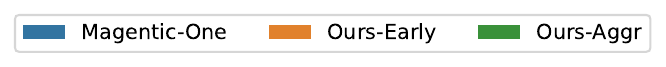}
        \includegraphics[width=\columnwidth]{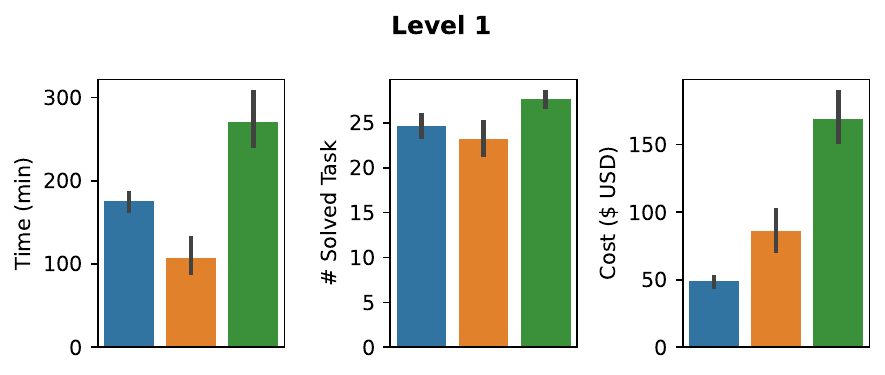}
        \includegraphics[width=\columnwidth]{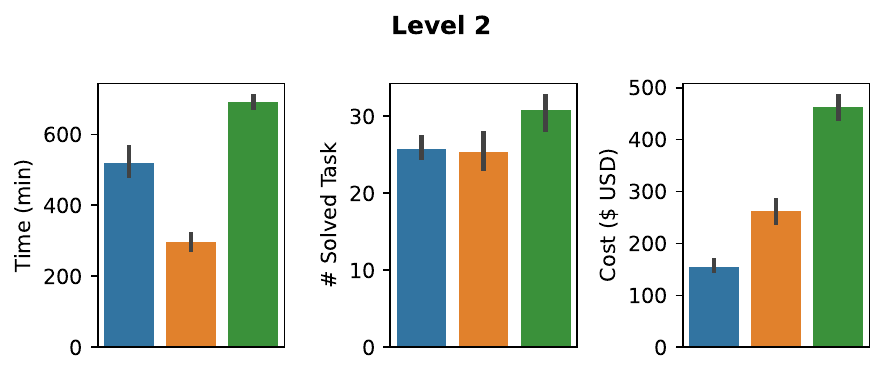}
        \includegraphics[width=\columnwidth]{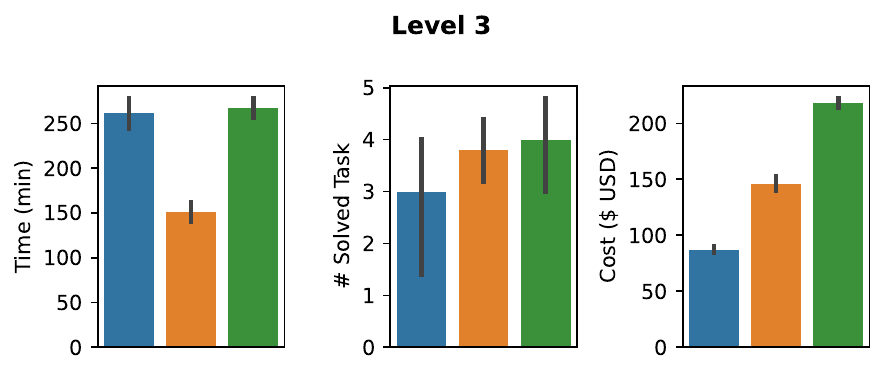}
        \caption{Latency, number of solved tasks, and monetary cost of different methods on GAIA. Error bars show the 95\% confidence interval.}
        \label{fig:gaia_main}
    \end{figure}
}

\newcommand{\GaiaVaryTeamsFigure}{
    \begin{figure}[t!]
        \centering
        \includegraphics[width=\columnwidth]{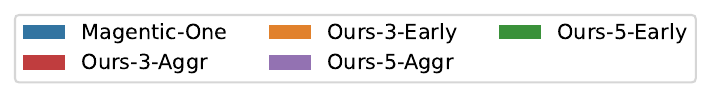}
        \includegraphics[width=\columnwidth]{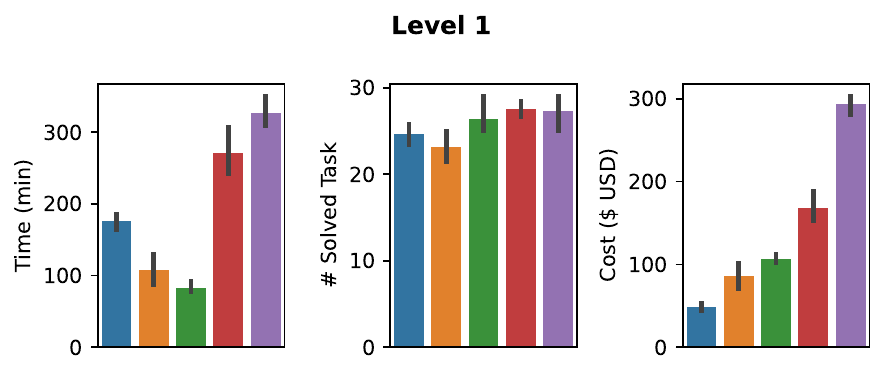}
        \includegraphics[width=\columnwidth]{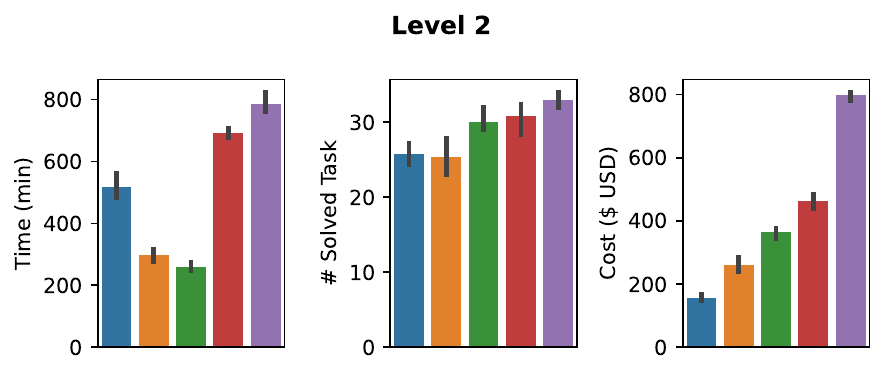}
        \includegraphics[width=\columnwidth]{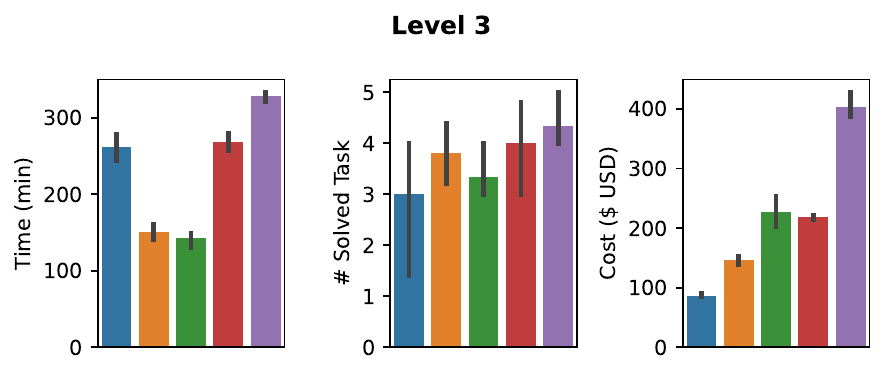}
        \caption{Latency, number of solved tasks, and monetary cost of Magentic-One and \system with varying number of teams. Error bars show the 95\% confidence interval.}
        \label{fig:gaia_vary_teams}
    \end{figure}
}

\newcommand{\GaiaAggrStrategyFigure}{
    \begin{figure}[t!]
        \centering
        \includegraphics[width=\columnwidth]{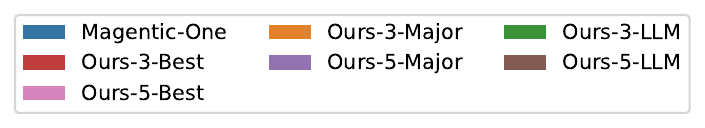}
        \includegraphics[width=\columnwidth]{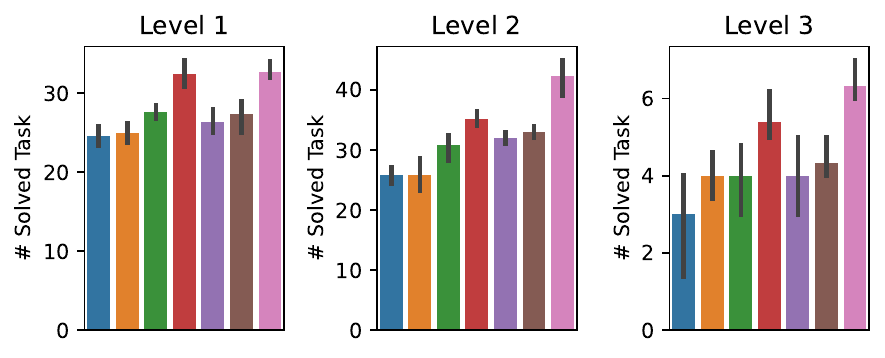}
        \caption{Number of solved tasks using Magentic-One and \system with different aggregation strategies and numbers of teams. Error bars show the 95\% confidence interval.}
        \label{fig:gaia_aggr_strategy}
    \end{figure}
}

\newcommand{\GaiaPlanStrategyFigure}{
    \begin{figure}[t!]
        \centering
        \includegraphics[width=\columnwidth]{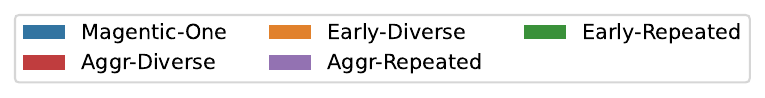}
        \includegraphics[width=\columnwidth]{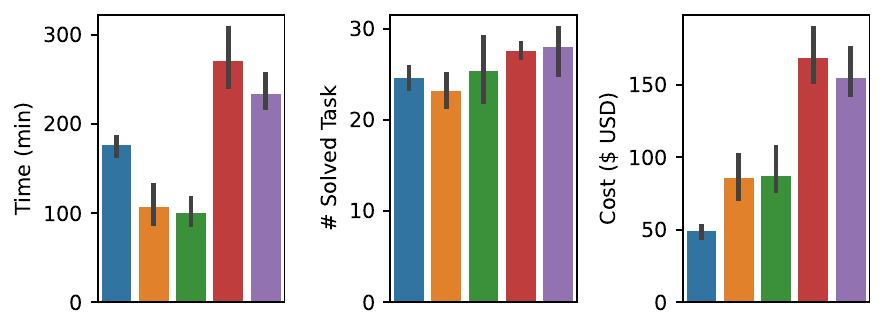}
        \caption{Latency of different planning strategies on GAIA level-1 tasks. Error bars show the 95\% confidence interval.}
        \label{fig:gaia_plan_strategy}
    \end{figure}
}

\newcommand{\GaiaSuccessSplitFigure}{
    \begin{figure}[t!]
        \centering
        \includegraphics[width=0.64\columnwidth]{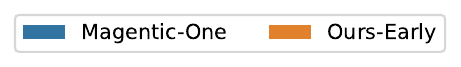}
        \includegraphics[width=\columnwidth]{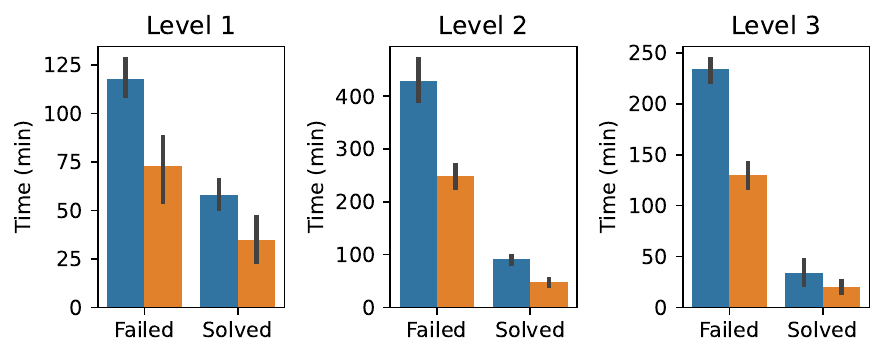}
        \caption{Latency breakdown for both solved and failed tasks on GAIA. Error bars show the 95\% confidence interval.}
        \label{fig:gaia_success_split}
    \end{figure}
}

\newcommand{\GaiaLatencyVarianceFigure}{
    \begin{figure*}[t!]
        \centering
        \includegraphics[height=12.5em]{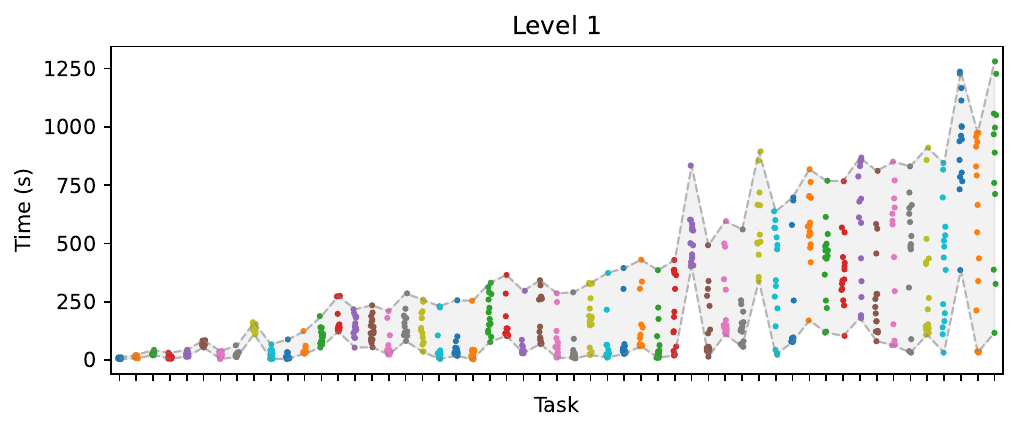}
        \includegraphics[height=12.5em]{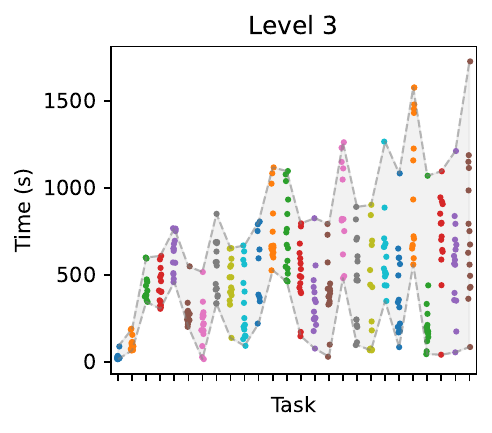}
        \includegraphics[height=12.5em]{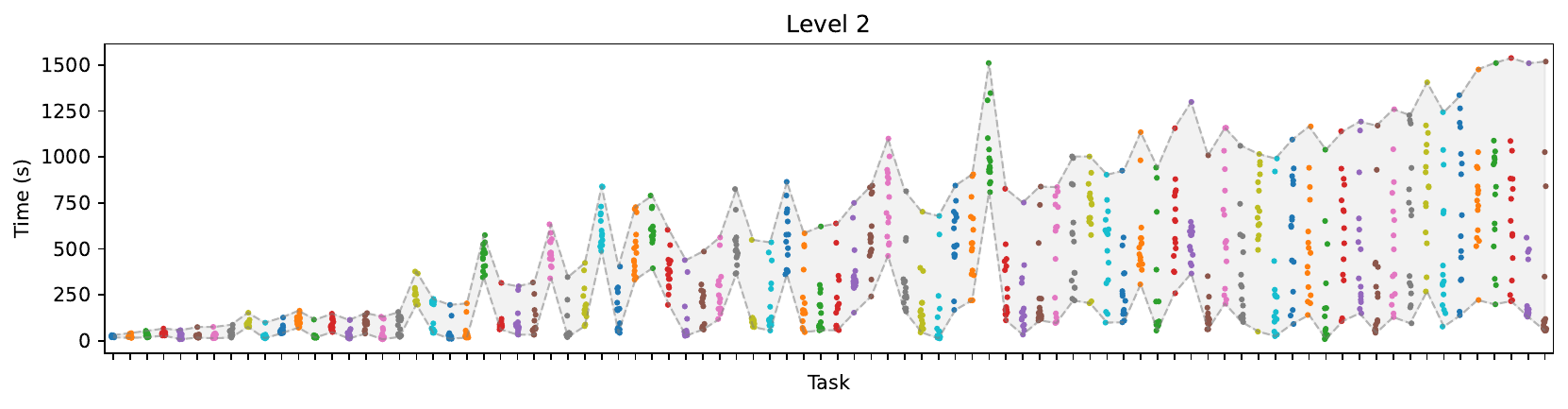}
        \caption{Latency distribution for \system teams on GAIA tasks, categorized by level. Each colored dot represents the time it takes a team to attempt a task. The x-axis shows the tasks sorted by their latency range (difference between highest and lowest latencies) in ascending order. The envelope and shaded region in gray span the minimum and maximum latencies for each task.
        Each task contains 15 data points (3 teams $\times$ 5 runs).}
        \label{fig:gaia_latency_variance}
    \end{figure*}
}

\newcommand{\GaiaLocalModelFigure}{
    \begin{figure}[t!]
        \centering
        \includegraphics[width=\columnwidth]{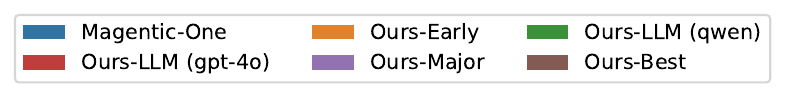}
        \includegraphics[width=\columnwidth]{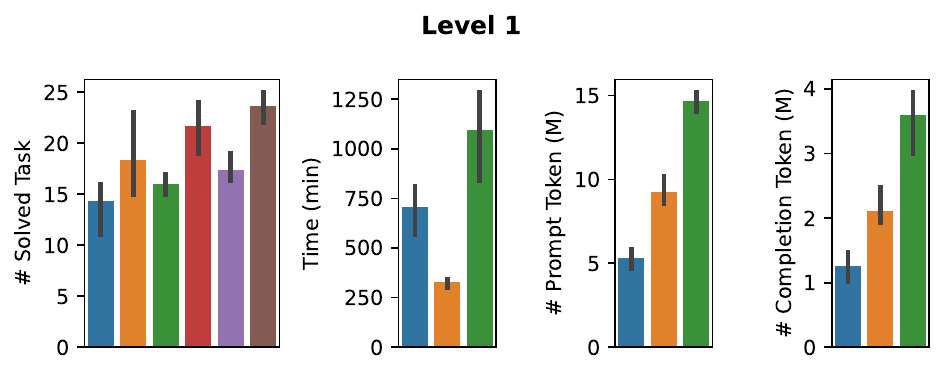}
        \includegraphics[width=\columnwidth]{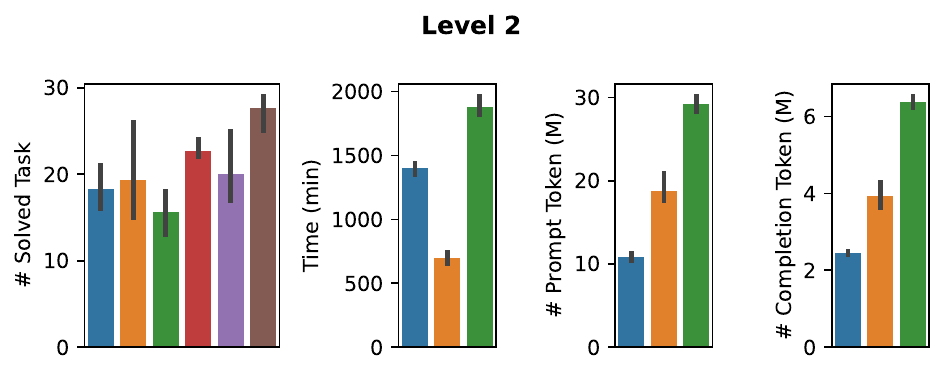}
        \includegraphics[width=\columnwidth]{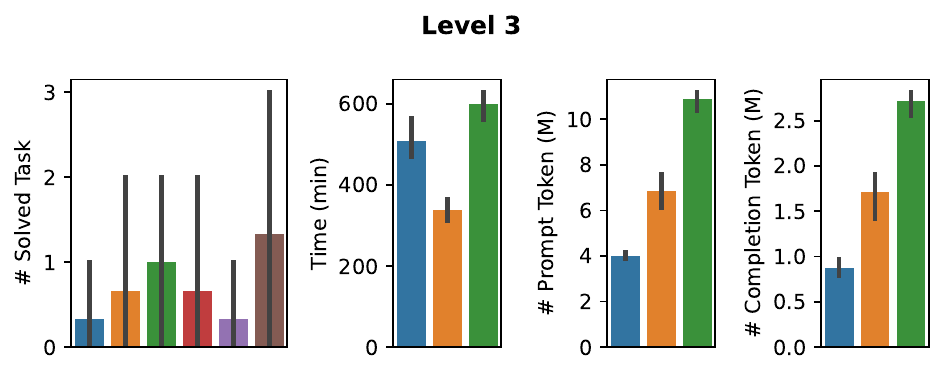}
        \caption{Number of solved tasks, latency, number of prompt tokens, and number of completion tokens with Qwen3-32B on GAIA. Error bars show the 95\% confidence interval.}
        \label{fig:gaia_local_model}
    \end{figure}
}
\newcommand{\teamStatsTable}{    
    \begin{table}[t!]
    \centering
    \caption{Mean latency (in seconds) and number of steps for \system on GAIA, grouped by the first (fastest), second, and third (slowest) teams that return an answer for each task. Darker shades of color indicate larger values.}
    \label{table:team_stats}
    {\renewcommand{\arraystretch}{1}
    \begin{tabular}{c|ccc|ccc}
        \hline
        \multirow{2}{*}{} & \multicolumn{3}{c|}{Latency (s)} & \multicolumn{3}{c}{Number of steps} \\ \cline{2-7}
         & 1st & 2nd & 3rd & 1st & 2nd & 3rd \\ \hline
        Level 1 
         & \cellcolor{red!12.5}{125} 
         & \cellcolor{red!20.2}{202} 
         & \cellcolor{red!30.1}{301}
         & \cellcolor{blue!10}{5} 
         & \cellcolor{blue!20}{8}  
         & \cellcolor{blue!33.3}{12} 
         \\
        Level 2 
         & \cellcolor{red!20.9}{209} 
         & \cellcolor{red!31.1}{311} 
         & \cellcolor{red!47.4}{474}
         & \cellcolor{blue!20}{8}  
         & \cellcolor{blue!30}{11} 
         & \cellcolor{blue!43.3}{15} 
         \\
        Level 3 
         & \cellcolor{red!36}{360} 
         & \cellcolor{red!47.8}{478} 
         & \cellcolor{red!62.2}{622}
         & \cellcolor{blue!36.7}{13} 
         & \cellcolor{blue!46.7}{16} 
         & \cellcolor{blue!60}{20} 
         \\ \hline
    \end{tabular}}
\end{table}
}
\newcommand{\aggregationPrompt}{
    \begin{figure}[t]
        \centering
        \begin{tcolorbox}[enhanced,
            colback=white,colframe=black,colbacktitle=gray,
            fontupper=\small]
            Earlier you were asked the following: \highlight{\{\textit{task}\}} \\
            Your team then worked diligently to address that request. You have been provided with a collection of transcripts and responses from \highlight{\{$n$\}} different teams to the question. Your task is to carefully evaluate the correctness of each team's response by analyzing their respective transcripts. After considering all perspectives, provide a FINAL ANSWER to the question. It is crucial to critically evaluate the information provided in these responses, recognizing that some of it may be biased or incorrect. \\
            Transcript from Team \#1: \highlight{\{$\textit{transcript}_1$\}}\\ 
            Response from Team \#1: \highlight{\{$\textit{response}_1$\}}\\
            $\cdots$ \\
            Transcript from Team \#n: \highlight{\{$\textit{transcript}_n$\}}\\ 
            Response from Team \#n: \highlight{\{$\textit{response}_n$\}}\\
            Let's think step-by-step. Carefully review the conversation above, critically evaluate the correctness of each team's response, and then output a FINAL ANSWER to the question.
        \end{tcolorbox}
        \caption{Aggregation prompt to integrate responses from multiple teams into a final answer.}~\label{fig:aggregation_prompt}
    \end{figure}
}

\newcommand{\planPrompt}{
    \begin{figure}[t]
        \centering
        \begin{tcolorbox}[enhanced, colback=white,colframe=black,colbacktitle=gray,
            fontupper=\small]
            Please devise another short bullet-point plan that is different from the above plans for addressing the original request. Try to consider different approaches to solve the task, different combinations of team members to use, and the sequnece in which to invoke them, etc.
        \end{tcolorbox}
        \caption{Diverse Plan Prompt.}~\label{fig:plan_generation_prompt}
    \end{figure}
}

\section{Introduction}
Large language models (LLMs) have shown strong reasoning capability across a variety of applications~\cite{DBLP:journals/corr/abs-2107-03374, 
NEURIPS2023_72223cc6,
DBLP:conf/nips/GuoGNLGCW023, DBLP:journals/pvldb/KayaliLFVOS24, 
DBLP:journals/corr/abs-2408-02243, DBLP:journals/tvcg/WangGBH25}.
Multi-agent systems with LLMs~\cite{wu23autogen,DBLP:journals/corr/abs-2406-04692,DBLP:conf/icml/Du00TM24,fourney2024magenticone} show further improvement in solving challenging reasoning tasks. A multi-agent system consists of a collection of agents that can interact and collaborate together to complete complex tasks. To achieve this, it requires agents of the system to decompose tasks into smaller steps, construct execution plans, use tools to complete tasks, communicate and exchange intermediate results, and dynamically revise plans when needed.

Many real-world tasks are \textbf{complex} and require multiple cycles of \textit{planning, acting, observing, and reflecting}~\cite{fourney2024magenticone}. Planning involves setting objectives and outlining the steps needed to achieve the objectives. Acting is the process of implementing each step of the plan and leveraging tools to enhance effectiveness. Observing refers to the collection of intermediate data on the implemented actions. Reflecting analyzes the observations and evaluates the effectiveness of the actions, which in turn informs adjustments for subsequent cycles.
Consequently, using LLM-based multi-agent systems for solving complex tasks often demands several iterative rounds of these stages, which can lead to high latency. 

As an example, a task from the GAIA benchmark~\cite{DBLP:conf/iclr/MialonF0LS24} is: ``How many studio albums were published by Mercedes Sosa between 2000 and 2009 (included)? You can use the latest 2022 version of english wikipedia.'' A state-of-the-art multi-agent system completes this in three minutes (including both the LLM inference time and additional delays such as waiting for websites to respond) over six steps. \Cref{fig:example_task} shows another GAIA task---this time, the same system spends 13 minutes across 20 steps attempting it but ultimately fails to arrive at the correct answer.

Existing methods reduce latency by parallelizing plans and executing independent subtasks in parallel~\cite{DBLP:conf/icml/KimMTLMKG24, DBLP:conf/iclr/Ning0ZWY024}, which have shown significant latency reduction for embarrassingly parallel workloads. However, many real-world complex tasks require step-by-step reasoning, where each step depends on the results of previous steps, thus are not suitable for plan parallelization. In addition, parallelization within a single plan is impractical for multi-agent systems~\cite{fourney2024magenticone} that dynamically devise plans and take actions at execution time.

\TaskFigure

We observe that there are often multiple correct plans to solve a task.
\Cref{fig:example_task} shows an example task from the GAIA benchmark~\cite{DBLP:conf/iclr/MialonF0LS24}, which requires multi-step reasoning to solve. A multi-agent system~\cite{fourney2024magenticone} can first search the birthplaces of U.S. presidents on Wikipedia, scroll down the page multiple times to collect the data, and then use internal knowledge of the LLM to identify the coordinates of the birthplaces (i.e., Plan \#1 in~\Cref{fig:example_task}). Alternatively, the system can visit the same Wikipedia page, but directly click and jump to the ``birthplaces'' section to find the data, and then perform multiple rounds of web search to find the coordinates (i.e., Plan \#2). In a third approach, it can visit the ``presidentusa.net'' website, click the ``view birthplaces'' button to collect the data, and then write a Python script using the ``geopy'' library to get coordinates of each city and run it to find the answer (i.e., Plan \#3). 
These plans are very diverse in terms of the number of steps and the types of agent invoked at every step. As a result, executing different plans can lead to very different latency. This opens the opportunity of reducing latency by concurrently executing multiple multi-agent system instances. 

Previous work has also shown that repeated sampling can improve reasoning performance of LLMs~\cite{DBLP:conf/iclr/ChenZNZLLC23, chen2024scalinglaws, DBLP:conf/icml/Du00TM24, DBLP:journals/corr/abs-2407-21787}, and various approaches to create diversity are explored to further improve performance~\cite{DBLP:conf/iclr/0002WSLCNCZ23, naik2024diversitythoughtimprovesreasoning, wang2024planningnaturallanguageimproves}. In this work, we explore the effect of sampling over multi-agent systems with actions to the task completion rate. We try different methods to encourage diversity and aggregate final answers from generated samples. However, our experiments indicate that diverse planning provides no clear advantage compared to repeated planning, likely due to the generation of suboptimal plans containing unnecessary steps intended solely to diversify the plans.

In this paper, we make the following contributions:
\begin{itemize}
    \item We show that parallel agents with early termination reduces latency without compromising the completion rate. We schedule multiple instances of multi-agent teams to solve tasks independently and concurrently, and terminating early whenever the first team completes the task.
    \item We show that parallel agents with aggregation improves the task completion rate, albeit at the expense of increased latency.  
    \item We explore different methods to encourage diverse execution plans; however, we did not observe significant performance improvements over repeated sampling.
\end{itemize}
\section{Related work}
\subsection{Latency optimizations in LLMs}

Many techniques have been proposed to accelerate LLM generation from the model or system perspective, including architecture design~\cite{DBLP:journals/jmlr/FedusZS22, DBLP:journals/corr/abs-2401-04088}, quantization~\cite{DBLP:conf/nips/DettmersPHZ23, DBLP:conf/nips/DettmersLBZ22}, sampling~\cite{DBLP:conf/icml/LeviathanKM23, DBLP:conf/nips/SternSU18}, and resource management~\cite{ye2025flashinfer, DBLP:conf/nips/DaoFERR22, DBLP:conf/sosp/KwonLZ0ZY0ZS23}. However, they require modifications to the models and the inference engines, which are infeasible in many applications. 

Another line of work optimizes latency at query time. LLMLingua~\cite{DBLP:conf/emnlp/JiangWLYQ23} compresses prompts, Batch Prompting~\cite{DBLP:conf/emnlp/ChengK023} batches multiple samples in one LLM invocation, Skeleton-of-Thought~\cite{DBLP:conf/iclr/Ning0ZWY024} first generates bullet points of the answer, then expands each point in parallel, and LLMCompiler~\cite{DBLP:conf/icml/KimMTLMKG24} decomposes user tasks into subtasks with inter-dependencies and executes them in parallel. These approaches either limit their applicability to certain types of simple tasks or require plans to be generated before execution. In contrast, \system optimizes latency for multi-agent systems that must dynamically generate plans during execution to tackle complex reasoning tasks.

\subsection{LLM-based multi-agent systems}

Multi-agent systems have shown strong potential to enhance the capabilities and performance of LLMs for complex problem-solving across various domains~\cite{DBLP:journals/corr/abs-2410-15364, qian-etal-2024-chatdev, DBLP:conf/ijcai/GuoCWCPCW024, DBLP:conf/acl/TangZ0L0ZCG24}. Methods such as Mixture-of-Agents~\cite{DBLP:journals/corr/abs-2406-04692} and Multiagent Debate~\cite{DBLP:conf/icml/Du00TM24} employ multiple agents to propose individual solutions and converge on a final answer through rounds of collective discussions. Other systems~\cite{qian-etal-2024-chatdev, fourney2024magenticone, DBLP:conf/naacl/WangMW0WJ24, DBLP:conf/iclr/HongZCZCWZWYLZR24} improve overall performance by configuring agents with different tools, roles, and personalities, thereby harnessing the diversity among agents to solve complex tasks step by step. While previous research primarily focused on improving task completion rates, \system also investigates from the system latency perspective. Moreover, it can be seamlessly integrated with existing multi-agent systems, offering an additional dimension for performance optimization.  

\subsection{Sampling in LLMs}
Repeated sampling has been demonstrated as an effective approach to improve LLM's capabilities. In code generation~\cite{DBLP:conf/iclr/ChenZNZLLC23, DBLP:journals/corr/abs-2107-03374, DBLP:journals/corr/abs-2203-07814}, performance can be improved by generating multiple candidates and verifying correctness via test cases. When verification tools are unavailable, techniques such as majority voting~\cite{DBLP:conf/iclr/0002WSLCNCZ23, DBLP:journals/corr/abs-2407-21787, chen2024scalinglaws}, user feedback~\cite{DBLP:journals/corr/abs-2208-05950, DBLP:journals/corr/abs-2408-02243}, proxy models~\cite{DBLP:journals/corr/abs-2110-14168, lambert2024rewardbench}, and answer aggregation~\cite{DBLP:conf/icml/Du00TM24} can be employed to derive a final answer. However, prior work has not examined how sampling can be used by multi-agent systems to address multi-step tasks. 
Moreover, prior work has focused on improving accuracy via sampling, while the effect of sampling on latency remains understudied. \system targets complex tasks that require multi-step reasoning in situations without verification tools and leverages parallel agents to reduce latency. 

Recent research also underscores the importance of maintaining diversity during sampling. Self-Consistency~\cite{DBLP:conf/iclr/0002WSLCNCZ23} encourages diversity through temperature and top-$k$ sampling. \textsc{Div-Se}~\cite{naik2024diversitythoughtimprovesreasoning} integrates diversity into the prompting strategy by explicitly asking the LLM to generate solutions using different approaches. Similarly, \textsc{PlanSearch}~\cite{wang2024planningnaturallanguageimproves} improves the performance of code generation by first generating diverse observations, followed by implementing codes derived from those observations. Building on these insights, we also explore techniques to encourage diversity, as reducing latency in \system relies on generating distinct execution plans with varying latencies. 
\section{Background: Magentic-One}
\system builds on Magentic-One. This section summarizes key background information about Magentic-One. 

Magentic-One is a generalist agentic system to solve complex reasoning tasks~\cite{fourney2024magenticone}. It consists of an \textbf{Orchestrator} agent along with four specialized agents: \textbf{WebSurfer}, \textbf{FileSurfer}, \textbf{Coder}, and \textbf{ComputerTerminal}.

The Orchestrator acts as a coordinator that directs specialized agents to achieve a high-level goal. Given a user task, the Orchestrator first creates an initial plan that decomposes the task into sub-tasks.
Following the plan, the Orchestrator invokes specialized agents in turn to solve sub-tasks. It tracks the progress to determine task completion and dynamically adjusts the plan based on the agent feedback.

The specialized agents are equipped with different tools and capabilities, including web interaction, file operation, code generation, and program execution. All agents except the ComputerTerminal are LLM-based.

Magentic-One also introduces a failure recovery mechanism, by tracking if the team is unable to make forward progress after a sequence of consecutive iterations. If a failure is detected, the Orchestrator will reflect on the failure and update a new plan to retry.

\section{Method}
\subsection{Parallel LLM agents}

\SystemDiagramFigure

We propose a new orchestration pattern that leverages parallel LLM agents. It consists of three steps: 

\textbf{(1) Plan generation:} Given a user task $q$, the centralized manager invokes a plan generation function $\pi$ to generate a set of $n$ plans, $P=\{p_i\}_{i=1}^n$:
$$P=\pi(q)$$
and instantiates a multi-agent team $f_i$ for each plan $p_i$. A multi-agent team could be any predefined pattern that can solve the user task. In our prototype, a team consists of an orchestrator, a coder, a computer terminal, a web surfer, and a file surfer~\cite{fourney2024magenticone}. 

\textbf{(2) Individual team completion:} Each team solves the task using the given plan, independently and concurrently. Once completed, it returns a binary success indicator $s_i\in\{0,1\}$,
an answer $a_i$ (only defined if $s_i=1$), along with a log $l_i$ that stores all the reasoning steps, to the centralized manager.
$$(s_i, a_i, l_i) = f_i(q, p_i)$$

To handle failure recovery, we also maintain a global memory module:
$$ M = \{(p, l) \mid \textrm{plan $p$ has failed with log $l$}\}$$

If the team fails to complete the task (i.e., $s_i=0$), the centralized manager first records the failure
$$M\leftarrow M \cup \{(p_i, l_i)\}$$
and then updates a new plan by leveraging real-time feedback from existing failures
$$p_i'=\pi(q, M)$$
and the team $f_i$ retries with the new plan $p_i'$. 

\textbf{(3) Early-stop or aggregation:} 
We study the benefits of parallel agents in terms of latency and task completion rate by proposing two modes (\Cref{fig:system_diagram}). For each team $f_i$, assume the user-perceived latency from the moment a task is submitted to the moment an answer is returned is $t_i$. We define $c_i(t_i)$ to be the cumulative monetary cost incurred by team $f_i$ up to time $t_i$. Denote the final latency, cost, and answer of \system as $t_s$, $c_s$, $a_s$. Define the sorted team latency in ascending order as
$$ t_{(1)} \leq t_{(2)} \leq \cdots \leq t_{(n)}$$

In the \textbf{Early-stop} mode (\Cref{subfig:early_stop}), the first answer returned by a team is treated as the final answer and the system is immediately terminated afterwards. Therefore,
$$ a_s = a_{(1)}, \quad t_s = t_{(1)}, \quad c_s = \sum_{i=1}^n c_i(t_{(1)})$$

In the \textbf{Aggregation} mode (\Cref{subfig:aggregation}), once $k$ answers are obtained, where $1 \leq k \leq n$, the centralized manager aggregates the collected information and generates a final answer.
$$ a_s = A(a_{(1)},\cdots , a_{(k)}), \quad t_s = t_{(k)}, \quad c_s = \sum_{i=1}^n c_i(t_{(k)})$$
where $A(\cdot)$ is an aggregation function that takes in a set of answers from multiple teams and returns a final answer. 

\subsection{Aggregation Strategy}~\label{subsec:aggregation}

\aggregationPrompt

We explore three instantiations of the aggregator $A(\cdot)$.

\textbf{Majority voting} (\texttt{major}) selects the most common answer.

\textbf{LLM-based} (\texttt{llm}) additionally incorporates the log of each team $l_i$ as input and prompts an LLM to aggregate a final answer: 
$$a_s=A(a_{(1)},\cdots , a_{(k)}, l_{(1)},\cdots , l_{(k)})$$ The aggregation prompt is listed in~\Cref{fig:aggregation_prompt}.

\textbf{Best-of-k} (\texttt{best}) measures if any answer $a_i$ from a team matches the ground truth. This method assumes the existence of an oracle aggregator and serves as the upper bound that our aggregation strategy can possibly achieve. 

\subsection{Diverse planning}

\planPrompt

The latency reduction relies on the ability of sampling different execution paths with varying latency. 
Literature has proposed various approaches to create diversity in LLM reasoning to improve performance~\cite{DBLP:conf/iclr/0002WSLCNCZ23, naik2024diversitythoughtimprovesreasoning, wang2024planningnaturallanguageimproves}. Therefore, we explore two different strategies to generate initial plans: 

\textbf{Repeated planning}. The centralized manager repeatedly and independently generates different plans using the same prompt with a high temperature value.

\textbf{Diverse planning}. We explicitly encourage diversity by asking the centralized manager to sequentially generate a new plan that is different from the existing ones. Given the user task $q$, the centralized manager first generates a plan: 
$$p_1=\pi_1(q)$$
Next, it generates the subsequent plans: $$p_i=\pi_2(q, p_1, \cdots, p_{i-1})$$
Here, $\pi_1$ and $\pi_2$ are plan generation functions, and we explicitly prompt the centralized manager to generate a different plan in $\pi_2$. The full prompts are listed in~\Cref{fig:plan_generation_prompt}.

However, we should note that these plans serve more like high-level thoughts for teams rather than plans that must be strictly followed.

\section{Experiments}

\noindent\textbf{Baselines.} We consider Magentic-One as our baseline, which is a representative multi-agent system. For our approach, we consider two settings: \system-Early is our parallel agents with early stopping. \system-Aggr is parallel agents with LLM-based aggregation. 

\noindent\textbf{Datasets.} We evaluate our approach on the \textbf{GAIA} dataset~\cite{DBLP:conf/iclr/MialonF0LS24}, which consists of realistic and challenging tasks that require various tool-use abilities and multiple steps of reasoning to solve. GAIA categorizes its tasks into three levels based on task difficulty, and splits tasks into a validation set and a test set. Since the answers of the test set are hidden, we evaluate on the validation set of each level. Specifically, there are 53 tasks in level 1, 86 tasks in level 2, and 26 tasks in level 3.

\noindent\textbf{Metrics.} We measure the system performance using latency, number of solved tasks, and monetary cost. For latency, we measure the wall-clock time as perceived by the user, starting when they submit a task and ending when the system returns the answer. To estimate monetary cost, we report the LLM inference cost using the OpenAI pricing model~\cite{openai_pricing}. Specifically, GPT-4o costs $\$2.50$ USD per 1M input tokens and $\$10.00$ USD per 1M output tokens. Both the latency and monetary cost are reported as aggregated values for running all tasks in the dataset.

\noindent\textbf{Evaluation setup.} 
We use the GPT-4o model (\texttt{gpt-4o-2024-08-06}) as the
underlying LLM for all agents given the complexity of the evaluated tasks. For the main experiment, each task is run five times. For other microbenchmarks, each task is run three times due to the high expense of LLM calls. By default, \system runs three team instances in parallel, using the LLM-based aggregation strategy and the diverse planning strategy. Following Magentic-One, a final prompt is used for all evaluated approaches to ensure the answer is expressed in a format expected by the benchmark dataset. \system is implemented in Python using the AutoGen 0.4 framework~\cite{autogen_github}. 

\subsection{Main results}~\label{subsec:exp_main}

\GaiaMainFigure

\Cref{fig:gaia_main} shows the latency, number of solved tasks, and monetary cost of each method across three levels on GAIA.
Compared with Magentic-One, \textbf{parallel agents with early stopping achieves $1.6\times$ to $1.8\times$ speedup, while maintaining the task completion rate.} In particular, level-3 tasks are challenging for all methods, resulting in a small number of solved tasks with high variance across methods. As a trade-off for executing multiple plans in parallel, \system also leads to a $1.7\times$ to $1.8\times$ increase in cost. However, the early stopping mechanism prevents it from reaching three times the baseline. 

Compared with Magentic-One, \textbf{parallel agents with aggregation improve task completion.} On average, \system solves three more tasks at level 1 (out of 53), five more tasks at level 2 (out of 86), and one more task at level 3 (out of 26). Because the execution time of \system is determined by its slowest team, it generally incurs greater latency than Magentic-One. The exception occurs at level 3, where the system struggles to solve the tasks and terminates due to reaching a maximum number of attempts. 
Furthermore, since \system launches three team instances, the monetary cost is increased by approximately threefold ($2.5\times$ to $3.4\times$).

\subsection{Varying number of teams}

\GaiaVaryTeamsFigure

To understand how the number of teams in \system affects the system performance, we vary the team number in \system. \Cref{fig:gaia_vary_teams} shows the performance of \system using both three teams and five teams on GAIA. As expected, \textbf{early stopping with five teams further reduces the latency, while maintaining the task completion rate}. \system with five teams obtains $1.8\times$ to $2.2\times$ speedup compared with Magentic-One, though the improvements are marginal compared with using three teams. Although the number of teams is increased to five, the cost increase is only $2.2\times$ to $2.6\times$ compared with Magentic-One. 

\textbf{Aggregation with more teams improves task completion as well.} Compared with Magentic-One, aggregation with five teams solves three, seven, and one more tasks in level 1, 2, and 3, respectively.

\subsection{Aggregation strategies}

\GaiaAggrStrategyFigure

\GaiaLocalModelFigure

\GaiaPlanStrategyFigure

\GaiaSuccessSplitFigure

\GaiaLatencyVarianceFigure

To evaluate how effective our LLM-based aggregation strategy (\texttt{LLM}) is, we compare against three other methods (as described in~\Cref{subsec:aggregation}): Magentic-One, which does not aggregate multiple solutions, majority voting (\texttt{Major}) and best-of-k (\texttt{Best}). \Cref{fig:gaia_aggr_strategy} shows the number of tasks solved by Magentic-One and \system under different aggregation strategies and varying team sizes. 
Our results show that \textbf{LLM-based aggregation is better than majority voting in many cases and does not harm performance in other cases}, suggesting that the logs of execution plans contain useful information that the LLM can leverage. 
However, a performance gap still remains between the LLM-based strategy and best-of-k. Currently, the log of each team is directly fed to the LLM to produce a final answer, but reasoning errors hidden in these logs can be difficult for the LLM to identify. Moreover, although the system supports multimodal task handling, the stored logs are text-only, which may further limit aggregation performance. These findings point to opportunities for further improvement in future work. 

\subsection{Local models}

We also investigate system performance when using local models. We test \system and the baseline on 
Qwen3-32B using one NVIDIA A100 GPU. \Cref{fig:gaia_local_model} shows the number of solved tasks, latency, number of prompt tokens, and number of completion tokens using Magentic-One and \system variants on GAIA tasks, including early stopping (\texttt{Early}), LLM-based aggregation using the same local model as the aggregator (\texttt{LLM (qwen)}), LLM-based aggregation using GPT-4o as the aggregator while using the local model in other system components (\texttt{LLM (gpt-4o)}), majority voting aggregation (\texttt{Major}), and best-of-k aggregation (\texttt{Best}). 
\textbf{With local models, parallel agents with early stopping maintain task completion rates, and parallel agents with aggregation using GPT-4o as aggregator improve task completion.}
Compared to the results shown in~\Cref{fig:gaia_main}, 
utilizing local models solve fewer tasks across all difficulty levels due to their smaller model sizes and reduced capabilities. Specifically, Qwen3-32B cannot effectively solve level-3 tasks, where all evaluated methods perform poorly. 
When using the same local model as aggregator, Qwen3-32B does not outperform Magentic-One. 
Specifically, 
Qwen3-32B tends to produce excessively long reasoning traces that confuse the model itself. 
However, replacing the local model with the more powerful GPT-4o as aggregator substantially improves task completion. Though this approach outperforms both majority voting and local model aggregation, a performance gap remains compared to best-of-k. Nevertheless, these results demonstrate that parallel agents with aggregation can improve task completion rate and highlight the potential for combining different models---such as using small models for individual teams and large models for aggregation---to optimize parallel agents.

Supporting parallel requests in local models requires more memory than our currently available hardware allows. While concurrent requests can be initiated, memory constraints force sequential queuing and processing. To evaluate potential performance gains under ideal conditions, we separately run a Magentic-One team three times and estimate the latency of \system with early termination as the minimum latency across the three runs and the latency of \system with aggregation as the maximum latency. 
As shown in~\Cref{fig:gaia_local_model}, \textbf{\system with early termination achieves $1.5\times$ to $2.2\times$ speedup, and also leads to a $1.6\times$ to $2.0\times$ increase in prompt and completion tokens as a tradeoff}. 
This shows that parallel agents with early termination could substantially reduce latency when sufficient hardware becomes available. 

\subsection{Planning strategies}

We next examine how different planning strategies impact system performance. \Cref{fig:gaia_plan_strategy} shows the performance of \system using repeated planning (\texttt{Repeated}) and diverse planning (\texttt{Diverse}) on GAIA level-1 tasks.
Interestingly, \textbf{diverse planning offers no clear advantage}. The results suggest that repeated planning even outperforms diverse planning by taking less time, completing more tasks, and requiring less cost. 
It is worth noting that in Magentic-One and \system, execution does not strictly adhere to the initial plan; instead, the orchestrator dynamically adjusts its plan based on actions and observations. In some cases, diverse planning generates suboptimal plans (e.g., introducing unnecessary steps to make the plan different), which could explain why it performs worse than repeated planning.

\subsection{Latency breakdown}

\Cref{subsec:exp_main} has shown that \system with early stopping brings a substantial reduction in overall latency. One might wonder whether this approach merely shortens latency for failed tasks without actually speeding up successful ones. 
\Cref{fig:gaia_success_split} addresses this concern by breaking down latencies for both solved and failed tasks. 
The results show that \textbf{parallel agents with early stopping lower latency for both solved tasks and failed tasks.} For tasks where solutions exist but can sometimes be slow to discover in a single-team scenario, other teams may reach a solution more quickly. For failed tasks that never achieve a correct solution, \system also helps reduce the ``long tail'' of particularly challenging paths by confirming infeasibility sooner.

To examine how different execution plans for the same task can result in different latencies, we visualize the latencies for all \system teams in our main experiment (3 teams $\times$ 5 runs). \Cref{fig:gaia_latency_variance} shows the latency distribution for the GAIA tasks, with the x-axis listing tasks in each level sorted by the time difference between highest and lowest latencies in ascending order. The results reveal substantial variance in latencies for each task across three levels. This spread arises from the diverse execution paths taken by each team. Notably, \system leverages exactly on this variance: by running multiple teams in parallel and terminating early when the first team finishes, the overall end-to-end latency is reduced. 

\teamStatsTable

Finally, \Cref{table:team_stats} shows the mean latency and number of reasoning steps required by \system to complete a task for the first (fastest), second, and third (slowest) teams. The table highlights clear performance differences among the teams, suggesting that faster teams also converge on solutions with fewer reasoning steps. Further, as the task becomes more challenging, there is a corresponding increase in both latency and the required number of steps.

\section{Conclusion}
In this paper, we present \system, a framework that optimizes multi-agent systems by executing multiple plans in parallel. We target complex reasoning tasks, which often have multiple valid solution paths with different latencies. Our experiments show that \system with early termination can reduce end-to-end latency without degrading task completion rates, while \system with aggregation can improve task completion rates at the expense of higher latency. We also investigate strategies for promoting plan diversity but observe no significant performance improvement compared to repeated sampling. Exploring other approaches to increase diversity can be an interesting future work for further system improvement.

\bibliography{paper}
\bibliographystyle{icml2025}

\end{document}